\documentclass[11pt,a4paper]{article}

\usepackage[utf8]{inputenc}
\usepackage[english]{babel}

\usepackage[top=20mm,bottom=30mm,left=20mm,right=20mm]{geometry}

\usepackage{authblk}
\usepackage{hyperref}       
\usepackage{url}            
\usepackage{amsfonts}      
\usepackage{nicefrac}       
\usepackage{microtype}      
\usepackage{amsmath}
\usepackage{amssymb}
\usepackage{physics}
\usepackage{graphicx}
\usepackage{tikz}
\usepackage[compat=1.1.0]{tikz-feynman}
\usepackage{subcaption}
\usepackage[export]{adjustbox}
\usepackage{wrapfig}
\usepackage{array}
\usepackage{cite}
\newcolumntype{M}[1]{>{\centering\arraybackslash}m{#1}}
\graphicspath{ {./images/} }

\author[1]{S. I. Godunov~\thanks{Corresponding author. E-mail: sigodunov@lebedev.ru}}
\author[1]{E. K. Karkaryan}
\author[1]{V. A. Novikov}
\author[2,3]{A. N. Rozanov}
\author[1]{M. I. Vysotsky}
\author[1]{E.~V.~Zhemchugov~\thanks{Corresponding author. E-mail: evgenii.zhemchugov@cern.ch}}

\affil[1]{I.E. Tamm Department of Theoretical Physics, Lebedev Physical Institute, 119991 Moscow, Russia}
\affil[2]{Institute for Theoretical and Experimental Physics, 117218 Moscow, Russia}
\affil[3]{Centre de Physique des Particules de Marseille, CPPM, Aix-Marseille
Universite, CNRS/IN2P3, F-13288 Marseille, France
}

\title{Forward proton scattering in association with muon pair production via the photon fusion mechanism at the LHC.}

\date{}
\begin{document}
\maketitle
\begin{abstract}
  Dilepton production in proton-proton collision through $\gamma\gamma$-fusion with one proton scattered elastically while the second produces a hadron jet is considered. Semi-analytical formulas describing the cross section of a muon pair production are presented.
\end{abstract}

Recent measurement of muon magnetic moment at Fermilab has confirmed the deviation from the Standard Model prediction~\cite{paper1}. When averaged with the previous BNL result~\cite{paper14}, it leads to a discrepancy of more than four standard deviations~\cite{paper1}. If it is a manifestation of New Physics, one should expect that at higher energies the deviations in the interactions of muons from the Standard Model predictions should be larger. Since ultraperipheral collisions are a source of very clean events, they can help to constrain parameters of new particles that can be responsible for this difference, for example, see~\cite{paper:dimuon, paper:gamma_X}.

At the Large Hadron Collider
(LHC), muon pairs are produced with high invariant masses, and this gives a chance for New Physics to be detected. It follows that the theoretical description of these reactions in the Standard Model is highly desirable. Recently the ATLAS collaboration has measured the fiducial cross section of the process when a muon pair is accompanied by the detection of one of the colliding protons in the forward detector~\cite{paper3}. Here we provide expressions for the cross section of this reaction. With help of the derived formulas, the cross section values can be evaluated by the standard numerical integration routines (for example, provided by the GSL~\cite{paper:gsl}) rather than Monte Carlo simulations. Let us note that numerical results for this reaction were recently presented in paper~\cite{paper4} (see also~\cite{paper5}).

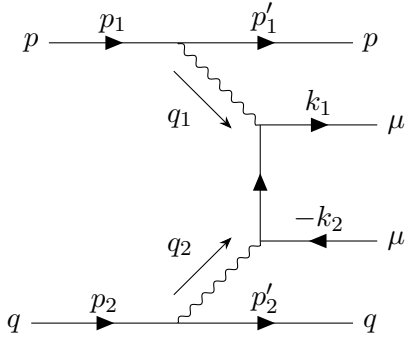
\begin{figure}[h]
\centering
\begin{tikzpicture}
  \begin{feynman}[]
  \vertex (a) {\(p\)};
  \vertex [right=5em of a] (s); 
  \vertex [right=6em of s] (c) {\(p\)};
  \vertex [below right=4em of s] (e);
  \vertex [right=4em of e] (g) {\(\mu\)};
  \vertex [below=4em of e] (f);
  \vertex [right=4em of f] (k) {\(\mu\)};
  \vertex [below left=4em of f] (m);
  \vertex [right=6em of m] (d) {\(q\)};
  \vertex [left=5em of m] (b) {\(q\)};
  \diagram*{
  (a) -- [fermion, edge label=\(p_1\)] (s) -- [fermion, edge label=\(p'_1\)] (c),
  (s) -- [photon, momentum'=\(q_1\)] (e) -- [fermion, edge label=\(k_1\)] (g),
  (f) -- [fermion] (e),
  (b) -- [fermion, edge label=\(p_2\)] (m) -- [fermion, edge label=\(p'_2\)] (d),
  (m) -- [photon, momentum=\(q_2\)] (f),
  (k) -- [fermion, edge label'=\(-k_2\)] (f),
  };
  \end{feynman}
\end{tikzpicture}
\caption{A $\gamma\gamma$ fusion mechanism of muon pair production.}\label{fig:diagram}
\end{figure}

\setlength{\parindent}{5ex}
The master formula which describes the reaction  under consideration can be easily obtained from the expressions provided in the review of two-photon particle production~\cite{paper6}:
\begin{equation} \label{eq1}
   d\sigma_{pq\rightarrow p\mu^+\mu^-q}=\frac{Q^2_q(4\pi\alpha)^2}{(q^2_1)^2(q^2_2)^2}(q^2_1\rho^{(1)}_{\mu\nu})(q^2_2\rho^{(2)}_{\alpha\beta})M_{\mu\alpha}M^*_{\nu\beta}\frac{(2\pi)^4\delta^{(4)}(q_1+q_2-k_1-k_2)d\Gamma}{4\sqrt{(p_1p_2)^2-m^4_p}} \frac{d^3p'_1}{(2\pi)^3 2E'_1} \frac{d^3p'_2}{(2\pi)^3 2E'_2},
\end{equation}
\noindent
where $\alpha$ is the fine structure constant, $Q_q$ is the electric
charge of the quark $q$, $\rho_{\mu \nu}^{(1)}$ and $\rho_{\alpha
\beta}^{(2)}$ are the density matrices of the photons, $M_{\mu\alpha}$ is the amplitude of the process, $p_1$, $p_1'$, $p_2$, $p_2'$ are the
proton and quark momenta before and after the collision (see Fig.~\ref{fig:diagram}),
$k_1$ and $k_2$ are the muons momenta, $q_1$ and $q_2$ are the photons momenta, $E_1'$, $E_2'$ are the proton and quark energies in the final state, $d \Gamma$ is the phase volume of the muon pair, and $m_p$ is the proton mass.
For the density matrices we get
\begin{equation}\label{eq2}
    \rho^{(1)}_{\mu\nu}=-\frac{1}{2q^2_1}Sp\{(\hat{p}'_1+m_p)\gamma_{\mu}(\hat{p}_1+m_p)\gamma_{\nu}\}=-\left( g_{\mu\nu}-\frac{q_{1\mu}q_{1\nu}}{q^2_1}\right)-\frac{(2p_1-q_1)_{\mu}(2p_1-q_1)_{\nu}}{q^2_1},
\end{equation}
and a similar expression for $\rho^{(2)}_{\alpha\beta}$. 

In order for the proton to remain intact, the square of the momentum transfer $-q^2_1$ should be bounded from above: $-q^2_1 \lesssim \hat{q}^2$, where $\hat q \sim \Lambda_{QCD}$. Following~\cite{paper7}, we take $\hat q = 200$~MeV in our calculations below. In this case, if the energy carried by the photon is in the interval $0.015E_p < \omega_1 < 0.15E_p$ then the proton will be detected by the forward detector with near $100\%$ efficiency~\cite{paper8, paper9}. As for the quark, its value of the transferred momentum $-q_2^2$ is
approximately bounded by the invariant mass of the muon pair $W$,
because for $-q_2^2 > W^2$ the cross section of the reaction $\gamma\gamma^* \to \mu^+ \mu^-$ quickly decreases as $W^2 /
(-q_2^2)$. Thus the photon with the momentum $q_1$ is emitted quasielastically and is polarized transversally, while the photon with the momentum $q_2$ can also be longitudinally polarized.\footnote{In order to suppress the background, only the events with the invariant
mass of the muon pair $W$ above a few GeV (e.g., 12~GeV~\cite{paper10}) are
selected. Therefore, neglecting corrections of the order of $\hat q^2 /W^2 \sim 3 \times 10^{-4}$, we can consider the photon with the momentum $q_1$ as real and polarized transversally.} 

\setlength{\parindent}{5ex}
The most appropriate way to deal with the density matrices $\rho^{(1)}_{\mu\nu}$ and $\rho^{(2)}_{\alpha\beta}$ is to introduce the basis of virtual photons helicity states. Let us suppose that in the center-of-mass system (c.m.s.) of the colliding photons we have $q_1=(\widetilde{\omega}_1, 0, 0, \widetilde{q}_1)$ and $q_2=(\widetilde{\omega}_2, 0, 0, -\widetilde{q}_1)$. The standard set of orthonormal four-vectors orthogonal to the momenta $q_1$, $q_2$ is
\begin{gather}
   e_\mu^{(1)}(+1) = \frac{1}{\sqrt{2}} (0, -1, -i, 0), \
   e_\mu^{(1)}(-1) = \frac{1}{\sqrt{2}} (0, 1, -i, 0), \
   e_\mu^{(1)}( 0) = \frac{i}{\sqrt{-q^2_1}} (\widetilde{q}_1, 0, 0, \widetilde{\omega}_1),
   \\ \nonumber 
   e_\mu^{(2)}(+1) = \frac{1}{\sqrt{2}} (0, 1, -i, 0), \
   e_\mu^{(2)}(-1) = \frac{1}{\sqrt{2}} (0, -1, -i, 0), \
   e_\mu^{(2)}( 0) = \frac{i}{\sqrt{-q_2^2}} (-\widetilde{q}_1, 0, 0, \widetilde{\omega}_2).
\end{gather}
These four-vectors correspond to the $\pm 1$ and $0$ helicity states of virtual photons in their c.m.s. They form a complete orthonormal basis for subspaces orthogonal to $q_{1\mu}$ and $q_{2\mu}$ respectively. Taking into account, that, due to the conservation of the vector currents, $\rho^{(1)}_{\mu\nu}q_{1\mu}=\rho^{(2)}_{\mu\nu}q_{2\mu}=0$, we obtain
\begin{gather}\label{eq4}
     \rho^{(i)}_{\mu\nu}=\sum_{a,b} [e^{(i)}_{\mu}(a)]^{*}e^{(i)}_{\nu}(b)\rho^{(i)}_{ab},\ a,b=\pm 1,0,
    \nonumber \\ \centering \rho^{(i)}_{ab}=(-1)^{a+b}e^{(i)}_{\mu}(a)[e^{(i)}_{\nu}(b)]^{*}\rho^{(i)}_{\mu\nu}.
\end{gather}
Here $\rho^{(i)}_{ab}$ are the density matrices in the helicity representation, and, according to \cite{paper11}, in the c.m.s. system of colliding protons in the case $E \gg \omega_1$, $xE \gg \omega_2$ we have~\footnote{In what follows, we consider high energy limit and neglect the masses of colliding particles.}
\begin{gather}\label{eq5}
    \rho^{(1)}_{++}=\rho^{(1)}_{--}=2\frac{E^2}{\omega^2_1}, \ \rho^{(1)}_{00}=4\frac{E^2}{\omega^2_1}, \nonumber \\
    \rho^{(2)}_{++}=\rho^{(2)}_{--}=2\frac{x^2E^2}{\omega^2_2}, \ \rho^{(2)}_{00}=4\frac{x^2E^2}{\omega^2_2},
\end{gather}
where $E\equiv\sqrt{s}/2=6.5 \ \text{TeV}$ is the colliding proton energy while $xE$ is the quark energy, $0<x<1$.

\setlength{\parindent}{5ex}
Finally we obtain 
\begin{gather}
    d\sigma_{pq\rightarrow  p\mu^+\mu^-q}=(4\pi\alpha)^2Q^2_q\frac{4(q_1q_2)}{4(p_1p_2)}\sigma_{\gamma\gamma^*\rightarrow\mu^+\mu^-}(q^2_2, W) \cdot 4\cdot 2 \bigg(\frac{E}{\omega_1}\bigg)^2 \cdot 2 \bigg(\frac{xE}{\omega_2}\bigg)^2 \frac{\frac{1}{2}dq^2_{1\perp} d\omega_1}{(2\pi)^4 q^2_{1}}\frac{\frac{1}{2}dq^2_{2\perp} d\omega_2}{(4xE^2) q^2_{2}}= 
    \notag 
    \\=\left( \frac{\alpha}{\pi} \right)^2 Q^2_q \frac{(q_1 q_2)}{(p_1 p_2)} \sigma_{\gamma \gamma^* \to \mu^+ \mu^-}(q_2^2, W) x E^2
   \frac{d q_{1\perp}^2}{q_{1}^2}
   \frac{d q_{2\perp}^2}{q_2^2}
   \frac{d \omega_1}{\omega^2_1}
   \frac{d \omega_2}{\omega^2_2},\label{eq6}
\end{gather}
where $(p_1p_2)=2E^2x$,\: $(q_1q_2)=(W^2-q^2_2)/2$,
\begin{gather}\label{eq7}
    d\sigma_{\gamma\gamma^*\rightarrow\mu^+\mu^-}=\frac{\sum\overline{|M|^2} \ d\cos{\theta}}{32\pi W^2(1-q^2_2/W^2)}, \nonumber \\
    \sum\overline{|M|^2}=\frac{1}{4}\big[|M_{++}|^2+|M_{+-}|^2+|M_{-+}|^2+|M_{--}|^2+2|M_{+0}|^2+2|M_{-0}|^2\big],
\end{gather}
$M_{\pm \pm}$, $M_{\pm 0}$ are the amplitudes of the process $\gamma\gamma^* \to \mu^+ \mu^-$ with the corresponding photons polarizations.
According to Eq. (E.3) from~\cite{paper6}, 
\begin{gather}\label{eq8}
    \sigma_{TT} \equiv \int \frac{1}{4}\big[ |M_{++}|^2+|M_{+-}|^2+|M_{-+}|^2+|M_{--}|^2 \big]\frac{d\cos{\theta}}{32\pi W^2(1-q^2_2/W^2)} \approx \nonumber \\ \approx 
    \frac{4\pi\alpha^2}{W^2}\Bigg[ \frac{1+q^4_2/W^4}{\big(1-q^2_2/W^2\big)^3}\ln{\frac{W^2}{m^2}} - \frac{\big(1+q^2_2/W^2\big)^2}{\big(1-q^2_2/W^2\big)^3} \Bigg], \nonumber \\
    \sigma_{TS} \equiv \int \frac{1}{2} \big[|M_{+0}|^2+|M_{-0}|^2\big]\frac{d\cos{\theta}}{32\pi W^2(1-q^2_2/W^2)} \approx \frac{16\pi\alpha^2W^2(-q^2_2)}{(W^2-q^2_2)^3},
\end{gather}
where $m$ is the muon mass. Thus, $\sigma_{\gamma\gamma^*\rightarrow\mu^+\mu^-}=\sigma_{TT} + \sigma_{TS}$ should be substituted in Eq.~\eqref{eq6}.

\setlength{\parindent}{5ex}
Integration over $q^2_{1\perp}$ in Eq.~\eqref{eq6} is easily performed: $\int dq^2_{1\perp}/q^2_{1} =2\ln{\hat{q}\gamma}/\omega_1$, where $\gamma = E/m_p$ is the Lorentz factor of the proton. In the c.m.s. of the protons, the following equations hold:
\begin{gather*}
    q_2=(\omega_2, q_{2\perp}, \frac{E_q}{p_q}\omega_2), \ q^2_2=-\frac{\omega^2_2}{x^2\gamma^2}-q^2_{2\perp}, \nonumber \\
    q_1q_2 =\omega_1\omega_2(1+\frac{E_q}{p_q})\approx 2\omega_1\omega_2, \ W^2=4\omega_1\omega_2+q^2_{2},
\end{gather*}
where $E_q$ and $p_q$ are the quark energy and spatial momentum: $p_2=(E_q,p_q)$.

\begin{figure}\label{fig}
\centering
\includegraphics[scale=0.9]{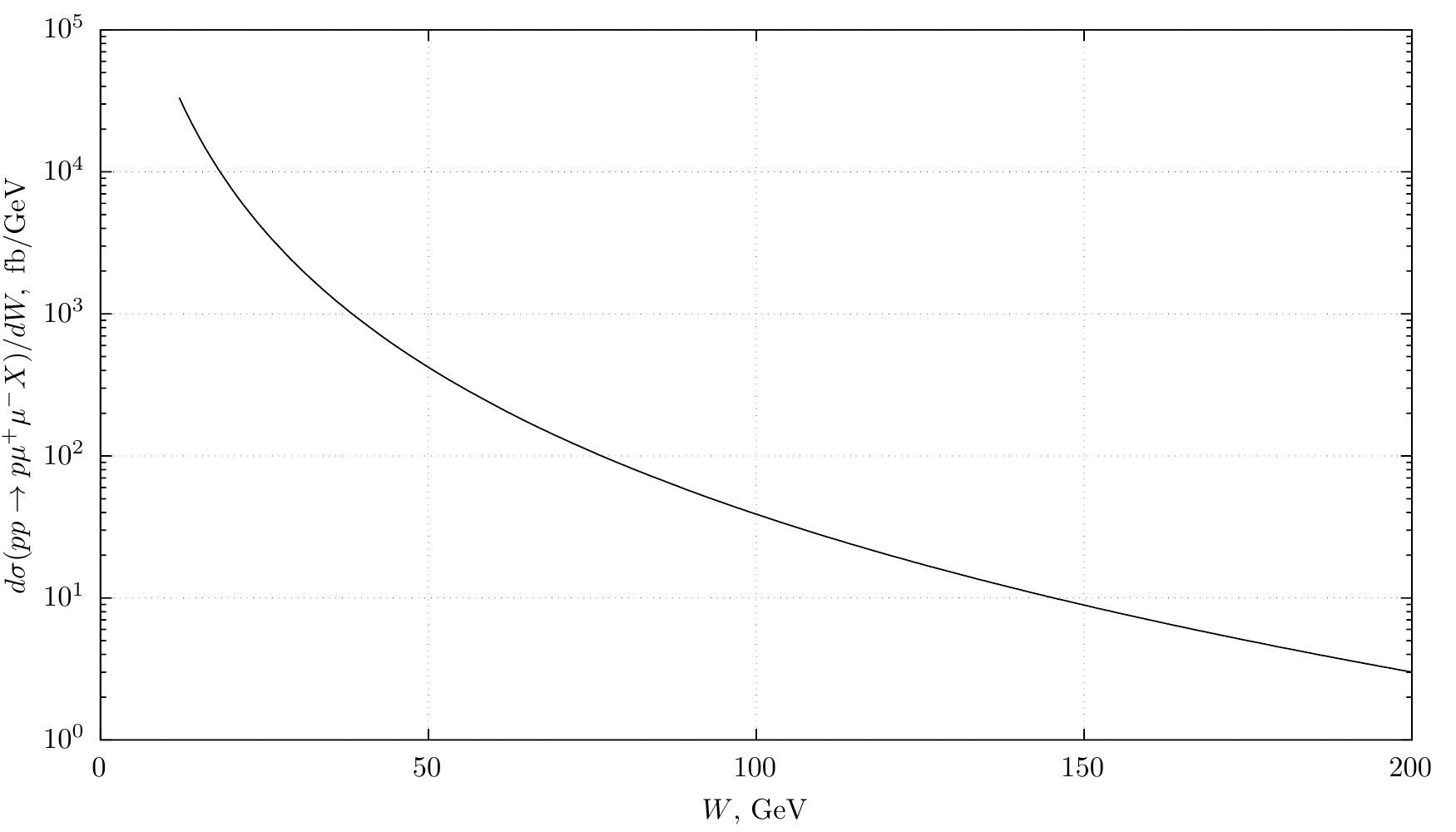}
\caption{The spectrum of muon pairs produced in photons fusion with one forward proton.}\label{fig:plot}
\end{figure}

It is convenient to change the integration variables from the photon energies $\omega_1$ and $\omega_2$ to the square of the invariant mass of the produced pair $W^2$ and the ratio of photon energies $y=\omega_1/\omega_2$: $d\omega_1d\omega_2dq^2_{2\perp}=(1/8y)dW^2dydQ^2_2$, where $Q^2_2=-q^2_2$. The following upper bounds on photon energies should be taken into account: $\omega_1 \leq \hat{q}\gamma$, $\omega_2 \leq xE$. Thus, we obtain
\begin{gather}
    \sigma_{pp\rightarrow p\mu^+\mu^-X}=2\cdot\Big(\frac{\alpha}{\pi}\Big)^2 \sum_{q} Q^2_q \int\limits^{\infty}_{\hat{W}^2}dW^2 \int\limits^{s\hat{q}/m_p-W^2}_0 \frac{\sigma_{\gamma\gamma^*\rightarrow\mu^+\mu^-}(Q^2_2, W^2)}{W^2+Q^2_2}dQ^2_2\int\limits^{1}_{\frac{W^2+Q^2_2}{s\hat{q}/m_p}}f_q(x, Q^2_2)dx\times\nonumber \\ \times \int\limits^{(2\hat{q}\gamma)^2/(W^2+Q^2_2)}_{(W^2+Q^2_2)/x^2s}\frac{dy}{y}\frac{\ln{(\hat{q}\gamma/\omega_1})}{Q^2_2+\big(\omega_2/x\gamma\big)^2}, \label{eq9}
\end{gather}
where $\omega_1=\sqrt{y(W^2+Q^2_2)}/2$, \ $\omega_2=\sqrt{W^2+Q^2_2}/(2\sqrt{y})$ and $f_q(x, Q^2_2)$ is the $q$-quark density function. Here the sum is taken over all quarks and antiquarks, both valent ($u$, $d$) and sea. The factor of $2$ takes into account the symmetrical process, when the proton from the second vertex remains intact. The differential cross section $d \sigma_{pp \to p \mu^+ \mu^- X}/dW$ for $W>12$~GeV is shown in Fig.~\ref{fig:plot}. 
Quark density functions were taken from \cite{paper12}. The integrated cross section in this region is
\begin{equation}\label{eq10}
    \sigma_{pp\rightarrow p\mu^+\mu^-X}(W>12\ \text{GeV})=203 \ \text{pb}.
\end{equation}
It is instructive to compare this result with the cross section of quasielastic $\mu^+\mu^-$ pair production \cite{paper7}: 
\begin{equation}\label{eq11}
    \sigma_{pp\rightarrow p\mu^+\mu^-p}(W>12\ \text{GeV})=60 \ \text{pb}.
\end{equation}

Let us mention the paper~\cite{paper13} in which the case of lepton pair production in photon fusion with both protons scattered inelastically is studied.

\section*{Conclusions}
The cross section of $\mu^+\mu^-$ pair production in semi-inclusive $pp$-scattering at the LHC is calculated (see Eqs.~\eqref{eq9} and \eqref{eq10}). The spectrum of the produced pairs is presented in Fig.~\ref{fig:plot}. Taking into account the dependence of differential cross section on photon virtuality explicitly, we have achieved better accuracy in comparison to the equivalent photon approximation.

We are supported by the Russian Science Foundation Grant No.  19-12-00123.

\end{document}